\documentclass[reprint,
superscriptaddress,
showpacs,
nofootinbib,
prd,
aps,
floatfix,
]{revtex4-1}

\usepackage{amsmath,amssymb,amsfonts,amsthm}
\usepackage[english]{babel}
\usepackage{url}
\usepackage{bm}
\usepackage[latin1]{inputenc}
\usepackage{graphicx,rotating}
\usepackage[colorlinks=true,linkcolor=black, citecolor=blue]{hyperref}
\usepackage{csvsimple}
\usepackage{booktabs}
\usepackage{multirow, array}
\usepackage{slashed}

\usepackage{natbib}

\begin{document}
\title{Detailed \texorpdfstring{$\beta$}{beta} spectrum calculations of \texorpdfstring{$^{214}$Pb}{214Pb} for new physics searches in liquid Xenon}
\author{L. Hayen}
\email[Corresponding author: ]{lmhayen@ncsu.edu}
\affiliation{Department of Physics, North Carolina State University, Raleigh, North Carolina 27695, USA}
\affiliation{Triangle Universities Nuclear Laboratory, Durham, North Carolina 27708, USA}

\author{S. Simonucci}
\affiliation{School of Science and Technology, Physics Division, University of Camerino, 62032 Camerino (MC), Italy}
\affiliation{INFN, Sezione di Perugia, 06123 Perugia (PG), Italy}
\author{S. Taioli}
\affiliation{European Centre for Theoretical Studies in Nuclear Physics and Related Areas (ECT*-FBK) \& Trento Institute for Fundamental Physics and Applications (TIFPA-INFN), Trento, Italy}
\affiliation{Institute of Physics, Nanotechnology and Telecommunications, Peter the Great St.Petersburg Polytechnic University, Russia}

\date{\today}
\begin{abstract}
We present a critical assessment of the calculation and uncertainty of the $^{214}$Pb $\to$ $^{214}$Bi ground state to ground state $\beta$ decay, the dominant source of background in liquid Xenon dark matter detectors, down to below 1 keV. We consider contributions from atomic exchange effects, nuclear structure and radiative corrections. For each of these, we find changes much larger than previously estimated uncertainties and discuss shortcomings of the original calculation. Specifically, through the use of a self-consistent Dirac-Hartree-Fock-Slater calculation, we find that the atomic exchange effect increases the predicted flux by $10(3)\%$ at 1 keV relative to previous exchange calculations. Further, using a shell model calculation of the nuclear structure contribution to the shape factor, we find a strong disagreement with the allowed shape factor and discuss several sources of uncertainty. In the 1-200 keV window, the predicted flux is up to 20$\%$ lower. Finally, we discuss omissions and detector effects in previously used QED radiative corrections, and find small changes in the slope at the $\gtrsim 1\%$ MeV$^{-1}$ level, up to $3\%$ in magnitude due to omissions in $\mathcal{O}(Z\alpha^2, Z^2\alpha^3)$ corrections and $3.5\%$ uncertainty from the neglect of as of yet unavailable higher-order contributions. Combined, these give rise to an increase of at least a factor 2 of the uncertainty in the 1-200 keV window. We comment on possible experimental schemes of measuring this and related transitions.
\end{abstract}

\maketitle

\section{Introduction}
With many experiments looking for Beyond Standard Model (BSM) physics signatures at the expected level of unavoidable background, an accurate calculation of the latter is crucial. In order to mitigate obvious sources of background, most of these experiments are located sufficiently deep underground so as to have a significant amount of overburden and effectively shield it from most cosmogenic radiation. As a consequence, the remaining background is typically related to naturally occurring radiation, either from the construction materials themselves or the surrounding underground laboratory environment. In the case of dual-phase liquid Xenon (LXe) detectors, the main backgrounds consist of isotopes of Xenon itself, and $\beta$ decay products of $^{220, 222}$Rn emanating into the detector volume. Dark matter or axion-like particle searches in this type of geometry are expected to show up in the keV range, placing stringent constraints on the calculation of $\beta$ spectra.

Recently, the XENON1T collaboration \cite{Aprile2020} observed an apparent excess of events relative to the background model in the very lowest energy bins, between 1-5 keV, for which several possible BSM possibilities were investigated and which have in turn inspired a flurry of phenomenological activity. At these low energies, the background is dominated by the ground state to ground state $\beta^-$ decay of $^{214}$Pb, which has a 27 minute half-life and is continually produced as a byproduct of the $^{222}$Rn $\alpha$ decay chain present in the environment and which emanates into the installation. The authors \cite{Aprile2020} stress the need for a precise calculation of the $^{214}$Pb $\beta$ spectrum near low energies for which a number of corrections were taken into account. Despite this, some of the approximations turn out not to be valid, which underscores both the difficulty and level of scrutiny required to make accurate predictions in a regime where scarcely any high-quality data exist, particularly for the mass range of interest. In the following sections we discuss a number of corrections which were either not included or calculated too crudely. We treat the atomic exchange correction in Sec. \ref{sec:exchange}, nuclear structure effects in Sec. \ref{sec:forbidden} and finally radiative corrections in Sec. \ref{sec:rc}. Section \ref{sec:summary} looks at the cumulative effect compared to the previous estimate and looks ahead. All the other components of the $\beta$ spectrum shape are based on a recent review \cite{Hayen2018} and its open-source implementation \cite{Hayen2019a}.

\section{Spectral corrections}
\subsection{Preliminaries}
\label{sec:prelims}
Practically all $\beta^-$ decays, while a weak quark-level process, occur not simply in a nucleus but typically inside an atom or molecule. As a consequence, the total Hamiltonian one must deal with is then
\begin{equation}
    \mathcal{H} = \mathcal{H}_\mathrm{nucl} + \mathcal{H}_\mathrm{e-e} + \mathcal{H}_\mathrm{weak},
    \label{eq:hamiltonian_general}
\end{equation}
where $\mathcal{H}_\mathrm{nucl}$ also contains the Coulomb potential of initial and final nuclear states\footnote{We have assumed here the Born-Oppenheimer approximation to separate the nuclear and electronic response.}. Because of the smallness of the weak interaction strength near zero momentum transfer appropriate to $\beta$ decay (the Fermi coupling constant is $G_F \approx 10^{-5}$ GeV$^{-2}$), it is sufficient for all practical purposes to treat it only to first order. Scattering states are eigenfunctions of the remaining part of Eq. (\ref{eq:hamiltonian_general}).

Since the energy transfer in nuclear $\beta$ decay is much smaller than the $W$ boson mass, we can up to zeroth-order in QED effects write the weak Hamiltonian as a local current-current interaction
\begin{equation}
    \mathcal{H}_\text{weak} = \frac{G_F}{\sqrt{2}}V_{ud}H_\mu L^\mu + \mathrm{h.c.}
    \label{eq:weak_hamiltonian_cc}
\end{equation}
where $V_{ud}$ is the up-down quark mixing matrix element, $L^\mu$ is the lepton current
\begin{equation}
    L^\mu = \bar{u}_e \gamma^\mu(1-\gamma^5) v_\nu
    \label{eq:L_mu}
\end{equation}
and $H_\mu$ a hadronic operator. While we hold off on a description of the latter until Sec. \ref{sec:forbidden}, Lorentz-invariance requires it to be a combination of vector, $V_\mu$, and axial vector, $A_\mu$, parts. Using this information we can write an initial Fock state
\begin{equation}
    | i \rangle = | J_i, M_i, T_i, T_{3i}, \alpha_i \rangle_\mathrm{nucl} | j_i, m_i; n_1^b \cdots n_k^b \rangle_\mathrm{e-e}
\end{equation}
where $J(T)$ denotes (iso)spin, $\alpha$ are additional quantum numbers and $n^b_i$ are bound electrons with suppressed quantum numbers combining to total angular momentum $j_i$. The final state can be constructed analogously with a continuum electron and antineutrino.

The $\beta$ spectrum shape can then be decomposed
\begin{align}
    \frac{d\Gamma}{dW_e} &= \frac{G_F^2 V_{ud}^2}{2\pi^3} F(W_e)C(W_e) R(W_e) X(W_e) K(W_e) \nonumber \\
    &\times p_eW_e(W_0-W_e)^2
    \label{eq:spectrum_shape}
\end{align}
where $F$ is the usual Fermi function which takes into account the Coulomb interaction between outgoing $\beta$ particle and the final nucleus, $X$ is the exchange correction discussed in Sec. \ref{sec:exchange}, $C$ is the so-called shape factor which takes into account the nuclear structure effects discussed in Sec. \ref{sec:forbidden}, $R$ are QED radiative corrections discussed further in Sec. \ref{sec:rc},  and $K$ is a collection of smaller corrections \cite{Hayen2018}. Equation (\ref{eq:spectrum_shape}) was written assuming $\hbar=c=m_e =1$ while $W_e = E_\mathrm{kin}/m_e + 1$ is the total $\beta$ particle energy, $W_0$ is its maximum value, and $p_e = \sqrt{W_e^2-1}$ is the $\beta$ momentum.

\subsection{Atomic exchange effect}
\label{sec:exchange}
From Eq. (\ref{eq:hamiltonian_general}) it is clear that initial and final electronic eigenstates are non-orthogonal due to the charge-change. This generates a richness as multiple decay channels open up. The most evident of these are atomic final state excitations (shake-up and shake-off), and while this process takes away small amounts of energy it does not appreciably change the spectrum shape at low energies \cite{Hayen2018}. In the case of $\beta^-$ decay, however, there is an additional possibility due to the indistinguishability of the emerging electron with the surrounding atomic electrons. Since the final state consist simply of a continuum electron, the $\beta$ electron can be thought to decay directly into a bound state in the daughter atom with the subsequent expulsion of the previously bound electron, i.e. an atomic exchange \cite{Bahcall1963}. This is a single-step process which interferes linearly with the tree-level $\beta$ decay amplitude and is found to increase the decay rate substantially near low energies \cite{Harston1992}. The reason for the latter is intuitively clear, as the probability for decay into a bound state and the creation of a continuum electron state depends on the overlap between the wave functions. As the $\beta$ energy decreases, the spatial extension of its wave function increases, and the overlap with bound state wave functions (with an extent on the order of the Bohr radius) increases.

The calculation of the exchange correction requires an accurate knowledge of both bound and continuum wave functions. It's well-known that orbital angular momentum, $\bm{L}$, doesn't commute with the Dirac Hamiltonian and the solution to the central equation contains a mixture of states categorized instead according to $K = \beta(\bm{\sigma}\cdot \bm{L} + 1)$, with eigenvalues $\kappa = -1 (+1)$ for $s_{1/2} (p_{1/2})$ states. The solution to the Dirac equation in a central potential is then commonly written as
\begin{equation}
    \phi_\kappa^\mu(\bm{r}) = \left(\begin{array}{c}
         \mathrm{sign}(\kappa)f_k(r) \chi_{-\kappa}^\mu(\hat r)\\
         g_\kappa(r)\chi^\mu_\kappa(\hat r)
    \end{array} \right)
\end{equation}
where $\chi^\mu_\kappa(\hat r)$ is a $2\times 1$ column matrix, and $g, f$ are purely radial functions.

In the easily tractable case for an allowed $\beta$ transition ($\Delta \pi = $ no, $\Delta J = 0,1$), the lepton field carries no orbital angular momentum and both leptons have total spin $j=1/2$. In this case the exchange effect under some approximations reduces to $X \equiv 1 + \eta$ where \cite{Harston1992}
\begin{equation}
    \eta = f_s(2T_s+T_s^2) + (1-f_s)(2T_p+T_p^2)
    \label{eq:eta_s_p_allowed}
\end{equation}
with $f_s$ the proportional effect of $s_{1/2}$ states
\begin{equation}
    f_s = \frac{g_{-1}^c(R)^2}{g_{-1}^c(R)^2+f_1^c(R)^2},
    \label{eq:f_s}
\end{equation}
where $\{g,f\}^c$ are continuum radial functions evaluated at the nuclear radius, $R$. Finally, 
\begin{equation}
    T_l = -\sum_{nl \in \gamma} \langle El| nl \rangle \frac{\{g,f\}_{n,\kappa_l}^b(R)}{\{g,f\}_{\kappa_l}^c(R)}
    \label{eq:exch_T}
\end{equation}
where $\{g, f\}$ is the large component radial wave function depending on $\kappa_l$, $n$ is the principal quantum number of the bound state, and $\langle El | nl \rangle$ represents the overlap between a continuum state of energy $E$ and bound state $| nl \rangle$. Finally, $\gamma$ is the electron configuration of occupied states, which corresponds to the parent configuration in the sudden approximation.

The calculations quoted in Ref. \cite{Aprile2020} are based on Eq. (\ref{eq:eta_s_p_allowed}) and use fits to numerically calculated atomic electric potentials by Salvat et al. \cite{Salvat1987} with a free fit parameter for every orbital to reach agreement with single-electron binding energy calculations. The second term in the rhs of Eq. (\ref{eq:eta_s_p_allowed}), i.e. $p_{1/2}$ contributions, was neglected together with higher $\kappa$ contributions for forbidden transitions. Of all the $|ns\rangle$ states, the exchange effect in the very lowest energy range predominantly occurs with the highest occupied $|ns\rangle$ states. As a consequence, the overlap integral in Eq. (\ref{eq:exch_T}) is extremely sensitive to the position of its $n-1$ radial nodes, which in turn depends on the precision of the wave functions. For the calculation of $^{214}$Pb performed in Ref. \cite{Aprile2020}, this poses a number of issues: ($i$) The fitted potentials of Ref. \cite{Salvat1987} are reported as a sum of three Yukawa functions over the entire range of the atom, and do not capture the oscillatory behaviour of the charge density; ($ii$) Only the electronic ground state was taken into account; ($iii$) Electron wave functions in Eq. (\ref{eq:f_s}) are evaluated only at the nuclear radius instead of integrated over the nuclear volume; ($iv$) In Eq. (\ref{eq:exch_T}) initial and final bound states are assumed orthonormal, i.e. $\langle n^\prime l^\prime | n l \rangle = \delta_{nn^\prime}\delta_{ll^\prime}$.

Further, high-quality $\beta$ spectroscopy data at very low energies are extremely scarce and theoretical calculations of Mougeot \textit{et al.} \cite{Mougeot2014} on which Ref. \cite{Aprile2020} is based reached agreement only through the introduction of a nonphysical screening correction \cite{Hayen2018} which enhanced the $\beta$ spectrum shape by $>5\%$ in the lowest energy range rather than diminish it by about $-1\%$. In Ref. \cite{Aprile2020} it is stated that the effect of including $p_{1/2}$ orbitals can make up for this difference. Using the same methods, however, this contribution was estimated to be $0.3\%$ at 1 keV \cite{Hayen2018}. To our knowledge, unfortunately no agreement from the same authors has been published. An independent study of $^{63}$Ni could also not reach agreement with measured spectra below $5$ keV \cite{Vanlangendonck2018}. It is hypothesized that a downturn in the exchange contribution of one or more orbitals near very low energies could be the cause, as this feature is not replicated in a more direct numerical calculation \cite{Morresi2018}. This downturn is also present in the exchange correction of Ref. \cite{Aprile2020}.

We have performed a direct numerical calculation which does not require the approximations specified above. Specifically, the electron density is calculated self-consistently in a Dirac-Hartree-Fock-Slater (DHFS) fashion \cite{Morresi2018}. The integration is performed numerically over the full nuclear volume, and takes into account all $\kappa$ values. The latter is a straightforward generalization and can, e.g., be found in Ref. \cite{Morresi2018}. We note that a further generalization of the latter using the Behrens-B\"uhring machinary in a molecular geometry is in preparation. Even so, the contributions of $|\kappa| > 1$ to the exchange effect can \textit{a priori} be estimated to be negligible for spin $0 \leftrightarrow 1$ transitions, since for $p_{3/2}$ orbitals \cite{Rose1961}
\begin{equation}
    g_{-2}^b(r)/g_{-1}^b(r) \approx \frac{r^{1+3(\alpha Z)^2/8}}{2^{[3-(\alpha Z)^2]/2}}
\end{equation}
in natural units, which evaluates to about $0.3\%$ at the nuclear radius so that its contribution is sub-percent even when the overlap integrals are of the same order as the $|ns \rangle$ orbitals. 

\begin{figure}[ht]
    \centering
    \includegraphics[width=0.48\textwidth]{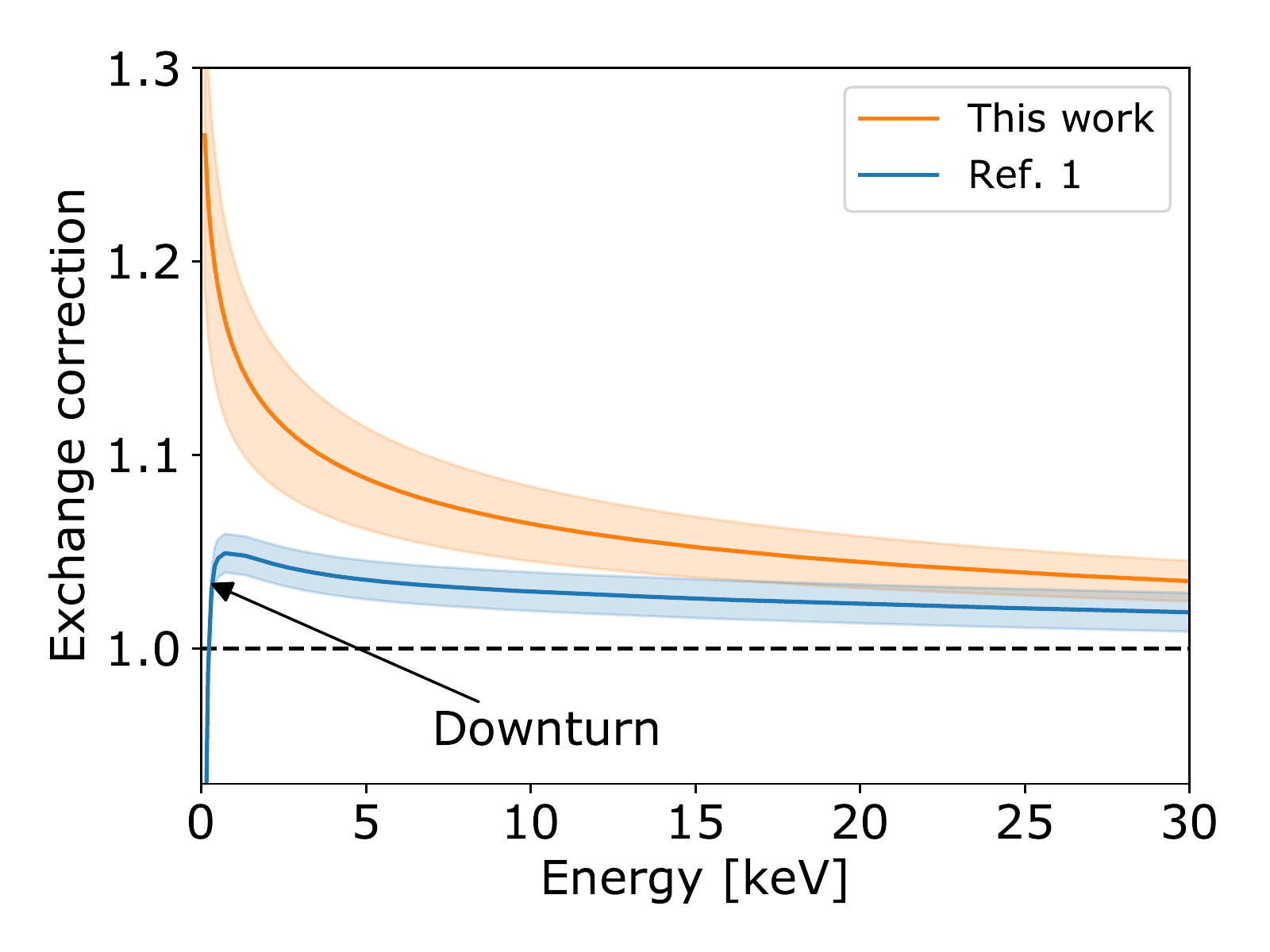}
    \caption{Comparison of our calculated exchange correction and that as reported by Ref. \cite{Aprile2020}. The latter was reported to have a flat $1\%$ uncertainty, shown in the blue band. We have chosen a $30\%$ relative error, shown in the orange band.}
    \label{fig:comparison_exchange}
\end{figure}

Our results for the exchange correction are shown in Fig. \ref{fig:comparison_exchange}, together with those reported in Ref. \cite{Aprile2020}. Our results point towards a substantially larger exchange effect in the very lowest energy range. Specifically, our exchange correction reaches $13\%$ at 2 keV, to be compared to $4\%$ from Ref. \cite{Aprile2020}, with the effect rising to $27\%$ at 0.1 keV. Even though the latter lies under the detection threshold of most experiments, a finite energy resolution will propagate some of this effect further. We note that we do not recover the downturn in our calculation. As this is a multiplicative term to the $\beta$ spectrum shape, the increase reported here translates directly into a decreased statistical significance of the reported XENON1T excess. We will come to back to this in Sec. \ref{sec:summary}.

Because of the scarcity of experimental data to compare to, it is hard to rigorously define a theoretical uncertainty. We remark that using the same methods, however, previously acquired theoretical results \cite{Morresi2018} show excellent agreement with the $^{63}$Ni and $^{241}$Pu spectra down to the lowest energy bins. Configuration interaction (i.e. beyond Hartree-Fock) estimates for lanthanum isotopes showed changes of $\mathcal{O}(5\%)$ near zero energy. As such, we make a conservative estimate of a $30\%$ relative error, and note that more data is needed to rigorously quantify it. This translates into a factor 4 larger uncertainty estimate compared to that of Ref. \cite{Aprile2020} at 2 keV.

\subsection{Forbidden shape factor}
\label{sec:forbidden}
We go into some more depth of the hadronic structure of the weak interaction Hamiltonian of Eq. (\ref{eq:weak_hamiltonian_cc}), the matrix element of which to zeroth order in QED is
\begin{align}
    _\mathrm{nucl}\langle J_f, M_f, T_f, T_{3f}, \alpha_f | &H_\mu | J_i, M_i, T_i, T_{3i}, \alpha_i \rangle_\mathrm{nucl} \nonumber \\
    \equiv \langle f| V_\mu &+ A_\mu | i \rangle
    \label{eq:strong_V_A}
\end{align}
where the notation is obvious. We use the Behrens-B\"uhring formalism, which consists of a spherical harmonics expansion of the currents and encodes all nuclear information into model-independent form factors \cite{Behrens1982}. As an example, the timelike component can then be written as 
\begin{align}
    H_0 = \sum_{LM}\mathcal{C}_{m_im_f;M}^{J_iJ_f;L}Y^M_L(\hat{q})\frac{(qR)^L}{(2l+1)!!}F_L(q^2),
    \label{eq:H_0_decomp}
\end{align}
where $\mathcal{C}$ denotes a Wigner-$3j$ symbol, $Y_M^L$ is a spherical harmonic, $q = p_f-p_i$ and $R$ is the nuclear radius so that $qR \ll 1$. All information is now encoded in the form factors $F_L(q^2)$, and if one performs a similar harmonic expansion of the leptonic current, Eq. (\ref{eq:L_mu}), the number of contributing form factors and their prefactors are singularly determined through angular momentum conservation. This is the so-called elementary particle method in a nutshell \cite{Behrens1982, Holstein1974, Armstrong1972a}. The shape factor can then be written for a non-unique first-forbidden $\beta$ decay as
\begin{align}
    C(W_e) &= M_1^2(1, 1) + m_1^2(1, 1) - \frac{2\gamma_1\mu_1}{W_e}M_1(1, 1)m_1(1, 1) \nonumber \\
    & + M_1^2(1, 2) + \lambda_2M_1^2(2, 1)
    \label{eq:C_explicit}
\end{align}
where the $\{M_1, m_1\}(\kappa_e, \kappa_\nu)$ are a linear combination of form factors, and $\kappa$ are the angular momentum eigenvalues from the lepton spherical wave expansion. Finally, $\gamma_1 = \sqrt{1-(\alpha Z)^2}$ and $\mu_1$ and $\lambda_2$ are $\mathcal{O}\{1+(\alpha Z)^2\}$ Coulomb functions \cite{Behrens1969}.

In practice, the form factors are reduced to calculable nuclear matrix elements through the impulse approximation. The latter assumes that the weak nuclear response can be written as a sum of responses of individual free nucleons moving in a mean-field potential. This allows one to write
\begin{equation}
    \langle f | O_K | i \rangle = \sum_{\alpha, \beta} \langle \beta | O_K | \alpha \rangle \langle f | [a^\dagger_\beta a_\alpha]^K | i \rangle 
\end{equation}
where $O_K$ is an order $K$ operator and $\alpha, \beta$ are single particle states. The first term is a single particle matrix element and can, e.g., be calculated analytically for spherical harmonic oscillator functions. The second term is a reduced one-body transition density (ROBTD), and is calculated using a many-body code such as the nuclear shell model.

The ground state to ground state decay of $^{214}$Pb [$0^+$] to $^{214}$Bi [$1^-$] is a so-called first-forbidden $\beta$ decay (${\Delta \pi = \mathrm{yes}}$, ${\Delta J = 0, 1, 2}$). Direct orbital filling in the standard Woods-Saxon potential leads to a proton ($\pi$) in the $\pi1h_{9/2}$ and neutrons ($\nu$) in the $\nu2g_{9/2}$ orbitals as suggested in Ref. \cite{Aprile2020}. The $1^-$ ground state of $^{214}$Bi, however, shows that even if this is the dominant configuration, an effective interaction must be present to arrive to a $1^-$ rather than $0^-$ ground state. The latter implies that the average occupation of a $(\nu 2g_{9/2} \pi 1h_{9/2})$ configuration is necessarily less than 1. Further, the matrix element $\langle 1h_{9/2} | O_{KLs} | 2g_{9/2} \rangle$ has only a single radial node from the initial state, so that the position of the node determines the partial cancellation between positive and negative contributions. In this respect, it is similar to the decay of $^{210}$Bi for which a large deviation in the shape factor is well-known \cite{Behrens1974a}. Unlike $^{210}$Bi, however, the partial half-life of the ground state to ground state $^{214}$Pb decay is not particularly hindered which would otherwise point towards significant cancellation.

Because of spin-parity requirements, only rank 1 operators can contribute to the decay, which to first order reduce down to three different non-relativistic  matrix elements to be calculated
\begin{equation}
    \mathcal{O}_1: g_V\bm{p}, \quad g_V\bm{r}, \quad g_A(\bm{\sigma} \times \bm{r})
\end{equation}
in the classic cartesian notation, with $g_{V (A)}$ the vector (axial vector) coupling constant. This is only the simplest picture, however, as the interaction of Eq. (\ref{eq:weak_hamiltonian_cc}) must be integrated over the full nuclear volume, which then folds in the spatial dependence of the leptonic wave functions. Once again then to first order the shape factor is determined by five form factors: $^VF_{101}, ^VF_{110}, ^AF_{111}, ^VF_{110}(1, 1, 1, 1)$ and $^AF_{111}(1, 1, 1, 1)$, where \cite{Behrens1970}
\begin{equation}
    F_{KLs}(\rho, m, n, k) \propto \int d^3r \phi_f^\dagger(r) O_{KLs}I(\rho, m, n, k;r)\phi_i(r)
\end{equation}
in impulse approximation, with $I$ a function originating from the expansion of the radial lepton wave functions. For the leading order terms this is
\begin{equation}
    I(1, 1, 1, 1;r) = \frac{3}{2}\left\{ \begin{array}{ll}
        1-\dfrac{1}{5}\left(\dfrac{r}{R}\right)^2 & \mathrm{for }~ 0 \leq r \leq R \\
        \dfrac{R}{r}-\dfrac{1}{5}\left(\dfrac{r}{R}\right)^3 & \mathrm{for }~ R \leq r 
    \end{array} \right.
\end{equation}
Finally, we note that the two vector form factors can be related to each other through the conserved vector current hypothesis, so that \cite{Behrens1978, Damgaard1966}
\begin{equation}
    ^VF_{101} = -\frac{1}{\sqrt{3}}\Delta_{T,T-1} R~ {}^VF_{110}
    \label{eq:CVC}
\end{equation}
where $\Delta_{T,T-1} \approx 23.85$ MeV is the excitation energy of isobaric analogoue state from which we extract $^VF_{101}$.

The full calculation taking into account all higher-order terms consists rather of 23 terms \cite{Behrens1974a}, even though the calculation is dominated by the first 5 we have mentioned above. We perform a shell model calculation with the NuShellX@MSU code \cite{Brown2014}, using the unconstrained \texttt{jj67pn} model space and the \texttt{khpe} interaction \cite{Warburton1991a}. The latter was used in particular for a study of $^{210}$Bi. Since this is the only interaction available for this region, our one-body transition densities are identical to those of another very recent study \cite{Haselschwardt2020}. We find good agreement with the experimental lifetime for an effective $g_A = 0.80 \pm 0.15$, in accordance with their results.

\begin{figure}[ht]
    \centering
    \includegraphics[width=0.48\textwidth]{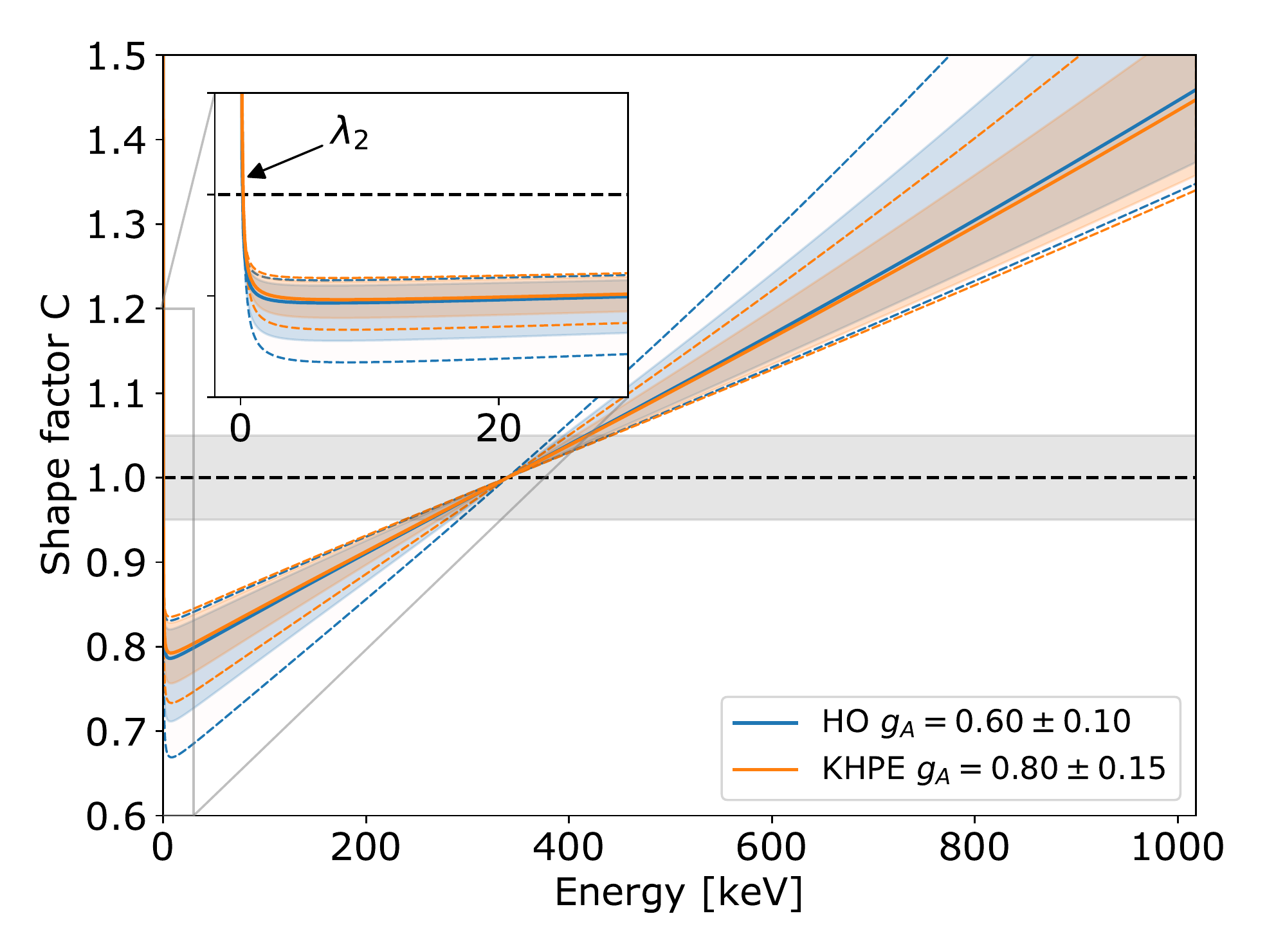}
    \caption{Calculated shape factor using the \texttt{khpe} shell model interaction using $g_A = 0.80 \pm 0.15$, for which we find almost identical results to Ref. \cite{Haselschwardt2020}. We show also the single particle estimate $\nu 2g_{9/2} \to \pi 1h_{9/2}$ for ${g_A = 0.60 \pm 0.10}$, and find good agreement as discussed in the text. The dashed lines correspond to changing the CVC prediction of Eq. (\ref{eq:CVC}) by 10\%. The inset shows the influence of the $\lambda_2$ Coulomb function, which becomes very large at low energies. All shape factors are normalized to unity at 1/3 of the endpoint energy for visual aid.}
    \label{fig:shape_factor}
\end{figure}

Our results are shown in Fig. \ref{fig:shape_factor}. We have additionally calculated the expected shape factor using a single particle transition $\nu 2g_{9/2} \to \pi 1h_{9/2}$, and find good agreement for the partial half-life using $g_A \approx 0.60$. This is not surprising since a more severe truncation of the model space generally requires stronger quenching to reach agreement. The shape factor, however, depends only the ratio of matrix elements and tends to be less sensitive to the specific value of $g_A$. In this case, since the shape factor is dominated by the ${}^AF_{111}$ form factor, the dependence on $g_A$ is small. We also show the resulting effect of changing Eq. (\ref{eq:CVC}) by 10\%, as this relation is only approximate and carries model dependence \cite{Damgaard1966, Behrens1982}. This has the effect of inflating the uncertainties by about $50\%$, and so has a reasonable influence on the shape factor. Finally, the \texttt{khpe} shape factor lies very close to harmonic oscillator estimate, which is reflected in the dominance of the $(\nu 2g_{9/2}\pi 1h_{9/2})$ configuration in the shell model ROBTDs, with only minor contributions from $\pi 2f_{7/2}$ and $\nu 1i_{11/2}, \nu 2g_{7/2}$ orbitals.

It is interesting, however, that at very low energies a significant enhancement occurs because of the $\lambda_2$ contribution in Eq. (\ref{eq:C_explicit}). In this regime, the $j=3/2$ component of the electron wave function in the presence of the final Coulomb potential is known to contribute significantly \cite{Behrens1969}. Throughout the rest of the spectrum this contribution is completely negligible. This appears to not be included in the results of Ref. \cite{Haselschwardt2020}.

The predicted spectrum amplitude is lowered by about $20\%$ in the 1-200 keV region of interest and a corresponding enhancement is found up to high energies. Within the context of the XENON1T experiment, this forces contributions from other sources of background to shift. This is particularly the case for the $^{133, 136}$Xe isotopes, for which the constraints on their activity are wide enough to accommodate such a shift. In the 1-30 keV range, however, the calculated shape factor is essentially flat and cannot influence the observed excess for a normalized spectrum, contrary to the claims of Ref. \cite{Haselschwardt2020}. The $\lambda_2$ contribution in our calculation is an exception here, however, as some of its effect will be observable through the energy resolution of the experiment.

\subsection{Radiative corrections}
\label{sec:rc}
Besides the Coulomb interaction between the emitted $\beta$ particle and the final state which is largely taken into account through the Fermi function, additional QED corrections contribute to $\mathcal{O}(\alpha)$ and beyond. The majority of this correction stems from real bremsstrahlung emission, with contributions from virtual photon exchange mainly responsible for removing the infrared divergence in a gauge-invariant way. These consist then of the well-known outer radiative corrections \cite{Sirlin1967}, which are largely model-independent. The requirements for which these expressions were derived, however, typically do not hold in calorimetric detection systems. Specifically, one assumes that the real photon that is emitted from the decay process goes either undetected or can be completely disentangled in energy from the $\beta$ particle. Up until some energy, however, this is typically not possible.

\begin{figure}[ht]
    \centering
    \includegraphics[width=0.48\textwidth]{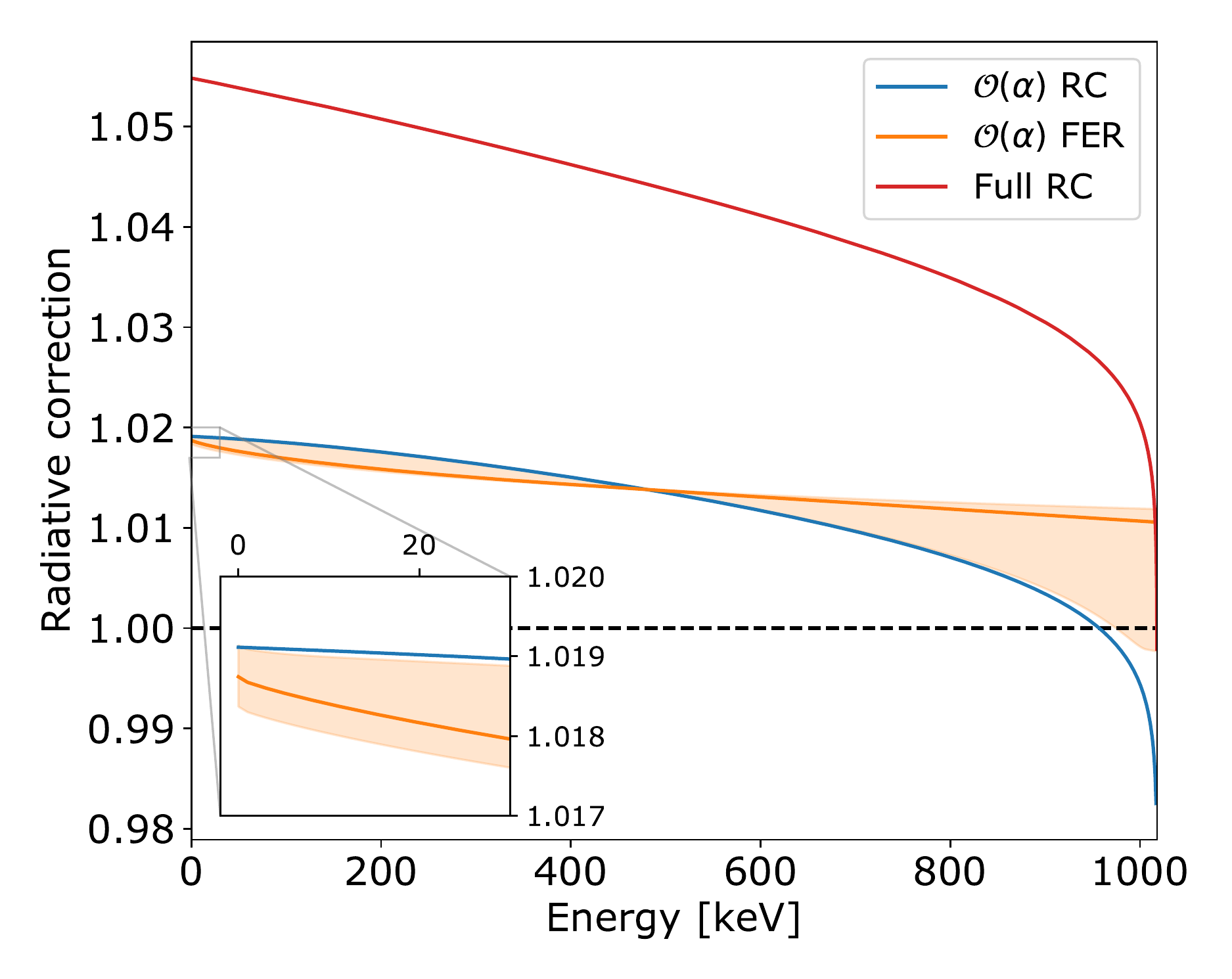}
    \caption{Comparison of $\mathcal{O}(\alpha)$ radiative corrections with a perfectly distinguishable photon, $\mathcal{O}(\alpha)$ RC using a finite energy resolution between 10 keV and infinity, and the full radiative corrections as in Ref. \cite{Hayen2018, Hayen2019a}, the latter assuming perfectly distinguishable photons.}
    \label{fig:rc_comp}
\end{figure}

In the case of LXe detectors, the path length of keV photons is on the order of micrometers, whereas the detector resolution is on the order of $1$ cm \cite{Aprile2019a}. As a consequence, all photons which cannot be distinguished from a $\beta$ particle are counted only in the total energy deposited. Below some threshold then, the sum $E_\mathrm{tot} = E_\beta + E_\gamma$ is counted, whereas above this threshold the analysis procedure decides what happens. There has been limited analytical work regarding this situation, and results are only available for $\mathcal{O}(\alpha)$ corrections \cite{Vogel1984, Kurylov2002, Gardner2004}. Figure \ref{fig:rc_comp} shows results for the standard $\mathcal{O}(\alpha)$ radiative corrections \cite{Sirlin1967}, the finite energy resolution (FER) results for a range of soft photon thresholds between 10 keV and infinity\footnote{We note for posterity that the last term in Eq. (9) of Ref. \cite{Gardner2004} should instead read $\Theta(E_e^\mathrm{max}-(\Delta E + E_e))\mathcal{I}(\Delta E, E_e)$.} \cite{Gardner2004}, and the full radiative corrections up to $\mathcal{O}(\alpha^3Z^2)$ \cite{Hayen2018, Hayen2019a} which we discuss below.

We make a crude estimate of a soft photon threshold energy from the position resolution of the XENON1T experiment \cite{Aprile2019} and X-ray stopping power tables \cite{Hubbell2004}. We take the conservative estimate that two vertices are distinguishable on average when separated by one full-width at half-maximum, and consider the effect after one absorption length. This crude estimate translates into a soft photon threshold of $\sim 400$ keV, which corresponds to the full FER line in Fig. \ref{fig:rc_comp}. The FER results are not very sensitive to the energy cut above a few tens of keV, with the results quickly approaching those of an infinite threshold. While a normalized spectrum will change the amplitude at 0 keV by up to a percent, the change in slope is only on the order of $0.3\%$ within a 0-200 keV window, but up to $3\%$ over the entire spectrum.

This brings us to higher order radiative corrections of $\mathcal{O}(Z^{n-1}\alpha^{n}) \equiv \delta_n$. The development of radiative corrections has a storied history and has been at the forefront of BSM physics searches for many decades \cite{Sirlin2013, Towner2008, Hayen2018}. Higher-order $\mathcal{O}(Z\alpha^2)$ and $\mathcal{O}(Z^2\alpha^3)$ terms are known within reasonable approximations, for which an exhaustive overview can be found in, e.g., Ref. \cite{Hayen2018}. The radiative corrections employed in Ref. \cite{Aprile2020} are energy-independent  by Mougeot \cite{Mougeot2015}
\begin{align}
    \delta_2^m &= 1.1 Z \alpha^2 \ln\left(\frac{m_p}{m_e}\right) \label{eq:delta_1_XM} \\
    \delta_3^m &= \frac{Z^2\alpha^3}{\pi}\left(3\ln 2 - \frac{3}{2} + \frac{\pi^2}{3} \right)\ln \left(\frac{m_p}{m_e}\right),
    \label{eq:delta_2_XM}
\end{align}
where $m_p$ is the proton mass and we have corrected what is assumed to be a typo for the leading log. These appear to be inspired by the works of Jaus and Rasche \cite{Jaus1970, Jaus1972}, although it is unclear how one arrived to, e.g., Eq. (\ref{eq:delta_1_XM}) from the original results in Ref. \cite{Jaus1970}. Regardless, throughout the decades following those results inconsistencies were resolved and missing terms identified \cite{Sirlin1986, Jaus1987}. The full results can be found elsewhere \cite{Hayen2018}. As an example we write the 
`model-independent' parts of $\delta_2$
\begin{equation}
    \delta_2^\mathrm{MI} = Z \alpha^2 \left[\ln\left(\frac{m_p}{m_e}\right) - \frac{5}{3}\ln(2W) + \frac{43}{18} \right]
\end{equation}
in the extremely relativistic approximation, while the model-dependent part requires knowledge of the charge distribution and is energy-independent. Similarly, $\delta_3$ has several energy-dependent terms, writing only the leading-log terms
\begin{equation}
    \delta_3(W) \approx Z^2\alpha^3 \biggl[6.63 \ln(W) -2.09 \ln^2(2W) - 0.65 \ln(2W_0)\biggr]
\end{equation}
where we have numerically evaluated the prefactors for convenience. This translates into the curve shown in Fig. \ref{fig:rc_comp}, which was calculated using the \texttt{BSG} library \cite{Hayen2019a} and where all terms were included. Relative to the order $\alpha$ results, this introduces a slope on the order of $1\%$ MeV$^{-1}$, and dominates in the absolute magnitude\footnote{We note that the proper radiative corrections also result in a 2.5\% lower magnitude than what is obtained from Eq. (\ref{eq:delta_1_XM}) and (\ref{eq:delta_2_XM}). This is not relevant for the spectral shape, but is directly proportional to the $ft$ value.}. Results up to this order have only been calculated for the leading-log terms, however, and an estimate of their uncertainty is not settled. Historically, a 100\% uncertainty was attached in the superallowed Fermi $ft$ analysis \cite{Towner2008}, which we adopt here. Additionally, no analytical results are available to take into account the soft photon threshold due to the computational complexity, so we simply conservatively estimate the uncertainty at $1\%$ from the behaviour of the $\mathcal{O}(\alpha)$ terms.

Finally, because $\alpha Z \approx 0.6$ for the $^{214}$Pb transition, even higher contributions become non-negligible. We are not aware of any calculations that have been performed, other than a heuristic estimate by Wilkinson \cite{Wilkinson1997}. The latter describes an estimated geometric series summation of $\delta_n$ for $n=4$ to infinity using only the leading log results \cite{Hayen2018}, which for $Z = 83$ results in a total contribution of $1.2\%$. These results are highly approximate, and not corroborated through more detailed calculations. As such, we attach a 200\% uncertainty to this value to allow for a sign change.

In summary, because of the contribution of corrected N$^{(2, 3)}$LO QED results, a finite soft photon threshold in calorimetric systems, and heuristic estimate of higher-order results, changes on the few percent level are obtained. If we take estimated uncertainties in quadrature, this results in an uncertainty of $3.5\%$ because of the extremely high $Z$ value and unknown detector response. In the original work \cite{Aprile2020} no uncertainty was included.

\section{Summary and outlook}
\label{sec:summary}
\begin{figure}[ht]
    \centering
    \includegraphics[width=0.48\textwidth]{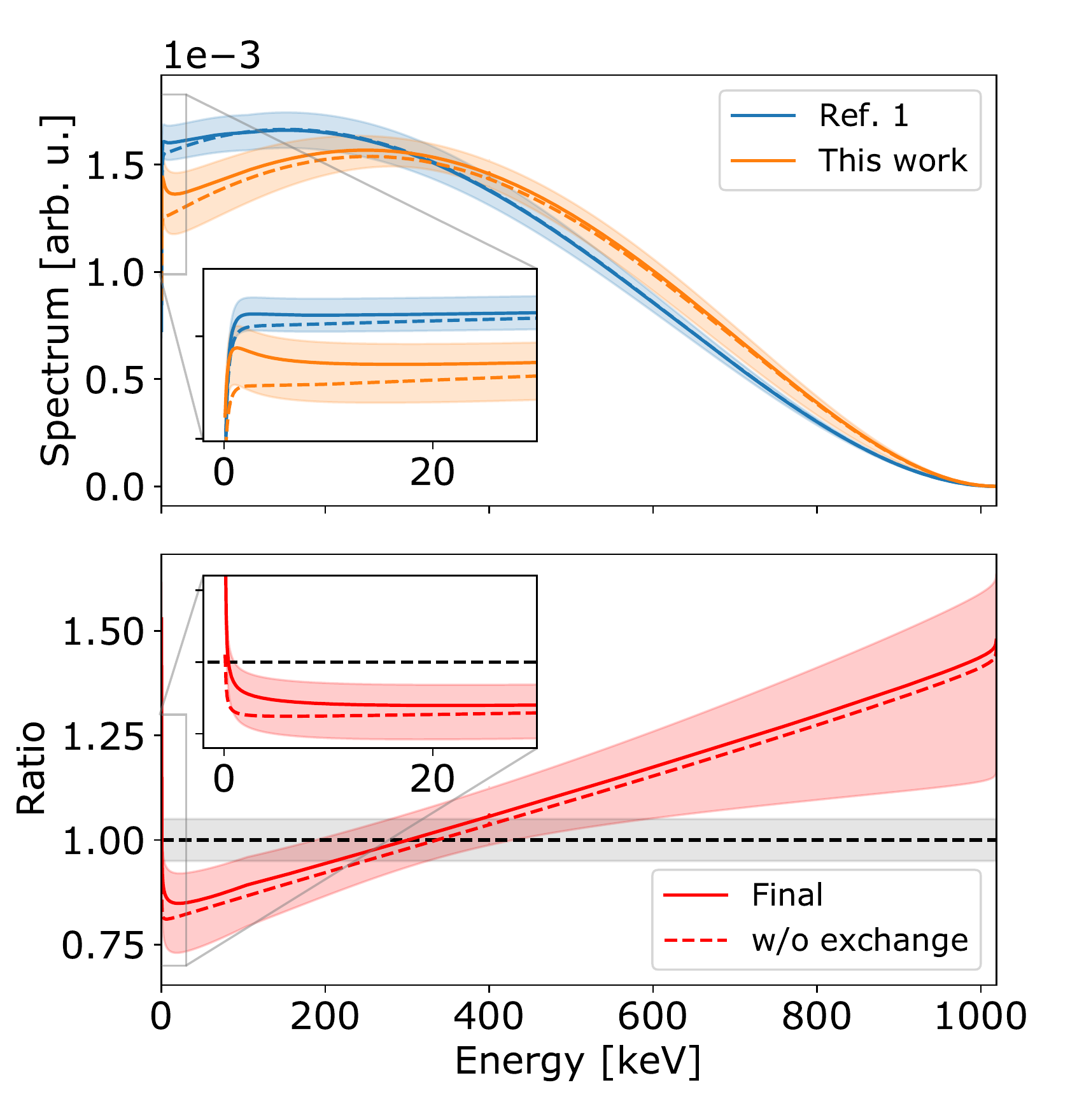}
    \caption{Overview of the spectral changes and uncertainties for the ground state to ground state $^{214}$Pb $\beta$ decay, the dominant background at low energies in LXe detectors. Top: difference in spectral shape between our work and Ref. \cite{Aprile2020}, with dashed lines showing results without the exchange effect. The common spectrum shape was calculated using \texttt{BSG} \cite{Hayen2019a} based on Ref. \cite{Hayen2018}. Bottom: Ratio between the spectral shapes. The gray band shows the quoted uncertainty of Ref. \cite{Aprile2020}, whereas the red band shows our uncertainty.}
    \label{fig:summary}
\end{figure}

An accurate description of $\beta$ spectra down to very low energies, i.e. the first few keV, is an extremely challenging task because of the multitude of effects that become prominent there. Since almost all of the latter are electromagnetic in origin, their importance and complexity scales roughly with $\alpha Z$. For LXe detectors whose main background at these energies arises from high-mass decay products of natural radioactivity, accurate calculations require heightened attention and scrutiny. We have commented here on three elements of the calculation of Ref. \cite{Aprile2020} for the ground state to ground state $^{214}$Pb to $^{214}$Bi $\beta$ decay and described significant improvements relative to their initial implementation. All of these give rise to changes much larger than the estimated uncertainty in Ref. \cite{Aprile2020}. Specifically, we have shown an increase of $>10(3)\%$ in the decay rate in the 1-3 keV regime, to be compared to the estimated uncertainty of $1\%$ of Ref. \cite{Aprile2020}. Further, detailed nuclear structure calculations show that the shape factor has a significant slope, leading to a change of almost $20\%$ in the 1-200 keV analysis and $60\%$ over the entire spectrum versus an estimated uncertainty of $5\%$ in Ref. \cite{Aprile2020}, confirming the results of \cite{Haselschwardt2020}. Interestingly, however, because of higher-order Coulomb effects not taken into account in Ref. \cite{Haselschwardt2020}, the shape factor contains an upward shift of more than $5\%$ in the 1-5 keV window. We have considered an additional source of uncertainty in the evaluation of an approximate form factor relationship dictated by CVC, which increases the shape factor uncertainty by an additional $50\%$. Finally, we have shown that the radiative corrections used in Ref. \cite{Aprile2020, Mougeot2015} are incomplete and are not consistent with the state-of-the-art at the few percent level in absolute magnitude and slope. Combined with estimates of detector efficiency and heuristic estimates of as of yet not calculated higher-order radiative corrections, an additional conservative uncertainty of $3.5\%$ was added. Our results are summarized in Fig. \ref{fig:summary}.

Significant changes occur throughout the entire spectrum and in the first few keV in particular. In the latter portion a steep change arises both from the $\lambda_2$ Coulomb function effect of Sec. \ref{sec:forbidden} and the significantly enhanced exchange effect of Sec. \ref{sec:exchange}. For the majority of the spectrum, the largest change is a result of a calculated shape factor with a positive slope significantly different from zero, discussed in Sec. \ref{sec:forbidden}. In the 1-200 keV region of interest of the XENON1T experiment \cite{Aprile2020}, this corresponds to a downward shift of almost 20\%. This implies a lower contribution of the other backgrounds, mainly above 100 keV. Here the background model is dominated by $^{133}$Xe and $^{136}$Xe, whose amplitude constraints can easily accomodate such a shift. Since the shape factor varies only very slowly on a keV scale, however, this large slope does not directly contribute to the observed excess, with the exception of our inclusion of the $\lambda_2$ contribution. Finally, we have estimated a $\beta$ spectrum shape uncertainty based on physical arguments which translates into at least a doubling of the uncertainty at low energies compared to Ref. \cite{Aprile2020}. As a consequence, the statistical significance of the observed excess will be reduced, although the final result depends on the complex analysis chain of the XENON1T experiment.

In the future, measurements of the $^{214}$Pb ground state to ground state $\beta$ spectrum and close neighbors will be vital in assessing the uncertainty budget of its inclusion, particularly because of its expected strong deviation from an allowed shape. Both the effects of the atomic exchange correction and the shape factor give rise to changes of several percents over modest energy ranges, although the lowest energies are extremely challenging to determine to very high precision. An interesting path forward in this endeavor is through the use of cyclotron radiation emission spectrometry (CRES). It is currently used for high precision measurements of the tritium endpoint energy by Project 8 \cite{Asner2015} and development work is being done to use CRES for, e.g., a full $^{6}$He $\beta$ spectrum at the University of Washington \cite{Graner2019} with the possibility of measuring gaseous isotopes up to the highest masses. This would provide invaluable and clean tests for the atomic calculations that go into the first tens of keV in $\beta$ decays throughout the nuclear chart. A measurement of the shape factor of $^{214}$Pb is particularly challenging because of the very similar lifetime of the $^{214}$Bi final state, and its production chain from $^{222}$Rn. A possibility that is being investigated uses an ion trap \cite{Li2013} to load purified $^{214}$Pb and measure its decay through $\beta$-$\gamma$ coincidence while using the $\alpha$-particle from the $^{214}$Bi chain as a tag.

\begin{acknowledgements}
One of the authors (L.H.) would like to thank Alejandro Garcia for the inspiration to look into this work, and Albert Young and Helena Almaz\'an for a careful reading of this manuscript. L.H. would additionally like to thank Albert Young, Alejandro Garcia and Guy Savard for stimulating discussions on experimental possibilities. L.H. acknowledges support from the U.S. National Science Foundation (PHY-1914133) and U.S. Department of Energy (DE-FG02-ER41042). S.S. and S.T. acknowledge the National Institute of Nuclear Physics (INFN) under the grant PANDORA.
\end{acknowledgements}

\bibliography{library}

\end{document}